\documentclass[aps,PRApplied,amsmath,amssymb,reprint]{revtex4-2}

\usepackage{graphicx}
\usepackage{dcolumn}
\usepackage{bm}

\usepackage[utf8]{inputenc}
\usepackage[T1]{fontenc}
\usepackage{array}
\usepackage{xspace}
\usepackage{booktabs}
\def\teod{$^{nat}$TeO$_2$\xspace}
\def\limo{Li$_2$$^{100}$MoO$_4$\xspace}
\def\NDBD{0$\nu\beta\beta$\xspace}
\def\DBD{2$\nu\beta\beta$\xspace}

\usepackage{lineno}
\usepackage{soul}
\usepackage{mathtools}
\usepackage{tabularx}
\usepackage{hyperref}
\hypersetup{
    colorlinks=true,
    linkcolor=blue,
    filecolor=blue,      
    urlcolor=blue,
    citecolor=blue,
    }

\begin{document}

\preprint{APS/123-QED}

\title[Large area photon calorimeter with Ir-Pt bilayer transition-edge sensor for the CUPID experiment]{Large area photon calorimeter with Ir-Pt bilayer transition-edge sensor for the CUPID experiment}

 \author{V.~Singh}
 \email{singhv@berkeley.edu}
    \affiliation{Department of Physics, University of California, Berkeley, Berkeley-94720, CA} 

\author{G.~Benato}
    \affiliation{GSSI-Gran Sasso Science Institute, I-67100, L'Aquila, Italy}  
    \affiliation{Department of Physics, University of California, Berkeley, Berkeley-94720, CA}

\author{M.~Beretta}
    \affiliation{Department of Physics, University of California, Berkeley, Berkeley-94720, CA}
 
 \author{C.~Capelli} 
    \affiliation{Nuclear Science Division, Lawrence Berkeley National Laboratory, Berkeley-94720, CA}
 
\author{C.L.~Chang} 
    \affiliation{High Energy Physics Division, Argonne National Laboratory, Lemont-60439, IL}
    \affiliation{ Kavli Institute for Cosmological Physics, University of Chicago, Chicago-60637, IL}

\author{B.K.~Fujikawa} 
    \affiliation{Nuclear Science Division, Lawrence Berkeley National Laboratory, Berkeley-94720, CA}

\author{E.V.~Hansen}
    \affiliation{Department of Physics, University of California, Berkeley, Berkeley-94720, CA}

 
\author{Yu.G.~Kolomensky}
    \affiliation{Department of Physics, University of California, Berkeley, Berkeley-94720, CA}
    \affiliation{Nuclear Science Division, Lawrence Berkeley National Laboratory, Berkeley-94720, CA}

\author{M.~Lisovenko}
    \affiliation{High Energy Physics Division, Argonne National Laboratory, Lemont-60439, IL}
    
\author{L.~Marini}
    \affiliation{INFN, Laboratori Nazionali del Gran Sasso, I-67100, Assergi, Italy} 
    \affiliation{Department of Physics, University of California, Berkeley, Berkeley-94720, CA}

 \author{WK.~Kwok}
    \affiliation{ Materials Science Division, Argonne National Laboratory, Lemont-60439, IL}
 
 \author{V.~Novosad}
    \affiliation{ Materials Science Division, Argonne National Laboratory, Lemont-60439, IL}
    \affiliation{Physics Division, Argonne National Laboratory, Lemont-60439, IL}

\author{J.~Pearson}
    \affiliation{ Materials Science Division, Argonne National Laboratory, Lemont-60439, IL}
    
\author{B.~Schmidt}
    \affiliation{Nuclear Science Division, Lawrence Berkeley National Laboratory, Berkeley-94720, CA}
 
\author{K.J.~Vetter}
    \affiliation{Department of Physics, University of California, Berkeley, Berkeley-94720, CA} 
    
\author{G.~Wang}
    \affiliation{High Energy Physics Division, Argonne National Laboratory, Lemont-60439, IL}
  
\author{B.~Welliver}
    \affiliation{Department of Physics, University of California, Berkeley, Berkeley-94720, CA} 
    \affiliation{Nuclear Science Division, Lawrence Berkeley National Laboratory, Berkeley-94720, CA}
    
\author{U.~Welp}
    \affiliation{ Materials Science Division, Argonne National Laboratory, Lemont-60439, IL}

\author{V.~Yefremenko}
    \affiliation{High Energy Physics Division, Argonne National Laboratory, Lemont-60439, IL}

\author{J.~Zhang}
     \affiliation{High Energy Physics Division, Argonne National Laboratory, Lemont-60439, IL}

\date{\today}

\begin{abstract}
CUPID is a next-generation neutrinoless double-beta decay experiment that will require cryogenic light detectors to improve background suppression, via the simultaneous readout of heat and light channels from its scintillating crystals. In this work we showcase light detectors based on a novel Ir-Pt bilayer transition edge sensor. We have performed a systematic study to improve the thermal coupling between the photon absorber and the sensor, and thereby its responsivity. Our first devices meet CUPID's baseline noise requirement of <100~eV rms. Our detectors have risetimes of $\sim$180~$\mu$s and measured timing jitter of <20~$\mu$s for the expected signal-to-noise at the Q-value of the decay, which achieves the CUPID's criterion of rejecting two-neutrino double-beta decay pileup events. The current work will inform the fabrication of future devices, culminating in the final TES design and a scaleable readout scheme for CUPID. 
\end{abstract}

\maketitle

\section{Introduction}
CUPID (CUORE Upgrade with Particle IDentification)~\cite{cupid_cdr} is the proposed next generation upgrade of CUORE (Cryogenic Underground Observatory for Rare Events)~\cite{adams2020improved} that will primarily search for the lepton number violating process of neutrinoless double beta (\NDBD) decay. The discovery of a lepton number violation would be a clear signature beyond the Standard Model of particle physics~\cite{particle2020review, dolinski2019neutrinoless}. Moreover, depending on the decay's underlying mechanism~\cite{dolinski2019neutrinoless,deppisch2012neutrinoless,rodejohann2012neutrinoless,prezeau2003neutrinoless,atre2009search,blennow2010neutrinoless,cirigliano2018neutrinoless}, the experiment can shed light on the neutrino's absolute mass scale, which is still unmeasured. CUPID will build on the experience of CUORE, an experiment that uses about a tonne of low temperature calorimeters operating close to $\sim$10~mK temperatures. CUPID will be housed in CUORE's state-of-the-art cryogenic infrastructure~\cite{alduino2019cuore} when the latter stops data collection. However, it will depart significantly from CUORE in its choice of detector. CUORE currently uses \teod crystals (50$\times$50$\times$50~mm$^3$) as low-temperature calorimeters, which are only sensitive to heat signals and therefore do not offer particle identification capabilities. In CUORE, the dominant background is from energy degraded $\alpha$ particle events in the crystal that can mimic the $\beta^{-}$ signals that we expect from the \NDBD process. CUPID aims to improve this using scintillating \limo (LMO) crystals (45$\times$45$\times$45~mm$^3$), where Mo will be enriched to contain > 95\% $^{100}$Mo. By simultaneously measuring thermal and scintillation signals, $\alpha$ events can be efficiently identified, as their light yield is a factor of five lower than that of $\beta$/$\gamma$ particles~\cite{armengaud2020cupid,poda2021scintillation}. 

If a \NDBD decay event occurs,  the two electrons emitted in the process deposit the transition energy of 3034~keV ($Q_{\beta\beta}$) in an LMO crystal. This results in a measurable heat signal and emission of sizeable scintillation light. In the CUPID detector, each LMO will be paired with a low-temperature auxiliary calorimeter with a wide area ($\sim$20~cm$^2$) to detect light with high photon collection efficiency. The energy of total light expected to hit an auxiliary calorimeter facing one of the LMO crystal surfaces is $\mathcal{O}(1~\text{keV})$ for $\sim$3~MeV electrons or gamma-rays~\cite{armatol2021characterization}. As a consequence, this device must have a very low energy threshold of $\mathcal{O}(100~\text{eV})$ rms. The detector should also have a good timing resolution to differentiate a pile-up event from two-neutrino double beta (\DBD) decays, which occur at a rate several orders of magnitude higher than \NDBD decay~\cite{chernyak2012random}. The slow response time of massive LMO calorimeters can cause an accidental pileup of \NDBD events and/or background events that will limit the sensitivity of the searches. The pileup effect can be severe for $^{100}$Mo, which is known to have a short half-life of 7.1$\times$10$^{18}$~yr amongst all \DBD decaying isotopes~\cite{armengaud2020precise}. The CUPID goal of keeping the \DBD decay pileup rate below 5$\times$10$^{-4}$~(counts/kg/keV/yr) requires an effective timing resolution of <170~$\mu$s on the light detectors~\cite{cupid_cdr}. The light detectors should also have a dynamic range of 10--15~keV to calibrate energy response with X-rays from Fe and Cu if needed. Other requirements include high radiopurity and low heat capacity of the materials used for the detector's construction. We have summarized the requirements of light detectors for CUPID baseline design in Table~\ref{tab:ldsummary}.

\begin{table}[t!]
\caption{\label{tab:ldsummary}Light detector requirements for the baseline design of the CUPID experiment.}
\begin{ruledtabular}
\begin{tabular}{cc}
Coverage area & 45~mm $\times$ 45~mm \\
Number of detectors & 1710 \\
Light absorption & > 90~\% \\ 
Baseline rms & < 100~eV \\
Pileup resolution & < 170~$\mu$s \\
Dynamic energy range & $\mathcal{O}$(10~eV) to $\mathcal{O}$(10~keV) \\
$^{238}$U contamination &  < 20 mBq/kg \\
$^{232}$Th contamination & < 10 mBq/kg \\
\end{tabular}
\end{ruledtabular}
\end{table}

Developing a high-resolution and high-efficiency detector coupled with a fast transition-edge sensor (TES) is promising. Sch\"affner et al.~\cite{schaffner2015particle} used silicon-on-sapphire discs to achieve a baseline resolution of $\sim$23~eV over a coverage area of $\sim$13~cm$^2$. Recently, Fink et al.~\cite{fink2021performance} demonstrated an excellent baseline energy resolution of $\sim$4~eV using Quasiparticle-trap-assisted Electrothermal feedback Transition-edge sensors (QETs)~\cite{irwin1995quasiparticle} on a surface area of $\sim$46~cm$^2$.  While both works used tungsten (W)-TES, there are inherent challenges in producing thin films at high throughput~\cite{abdelhameed2020deposition}. TES production requires a highly controlled environment with reproducible crystallinity and minimal structural and radioactive impurities in thin films. Furthermore, since the transition temperature is an inherent property of the material, it is difficult to tune the transition temperature ($T_c$) of W-TES by controlling its crystalline structure or introducing ferromagnetic impurities.  Tuning is critical for applications where lowering $T_c$ can improve energy resolution and better match the operating temperature of the cryogenic apparatus.  

In our previous work~\cite{hennings2020controlling}, we have detailed the fabrication recipe for producing Ir-based films with predictable and reproducible critical temperatures. We used a bilayer TES where we altered the $T_c$ of a superconducting Ir film using a normal Pt film on top of it. We use the superconducting proximity effect~\cite{meissner1960superconductivity,de1964boundary} to tune the $T_c$ by varying the thickness of the Pt film. The thermal characteristics of the Ir-Pt bilayer have been presented in Ref.~[ \onlinecite{zhang2022characterization}] where we reported the electron-phonon (e-ph) coupling strengths for the Ir-Pt bilayer and demonstrated that the addition of normal metal gold pads overlapping the TES and the photoabsorber improves the thermal coupling between them.

In this paper, we present one of the first light detector characterizations using Ir-Pt bilayers and showcase their potential as a suitable detector technology for CUPID.  

\section{Detectors} 
\begin{figure}[h]
\includegraphics[width=\linewidth]{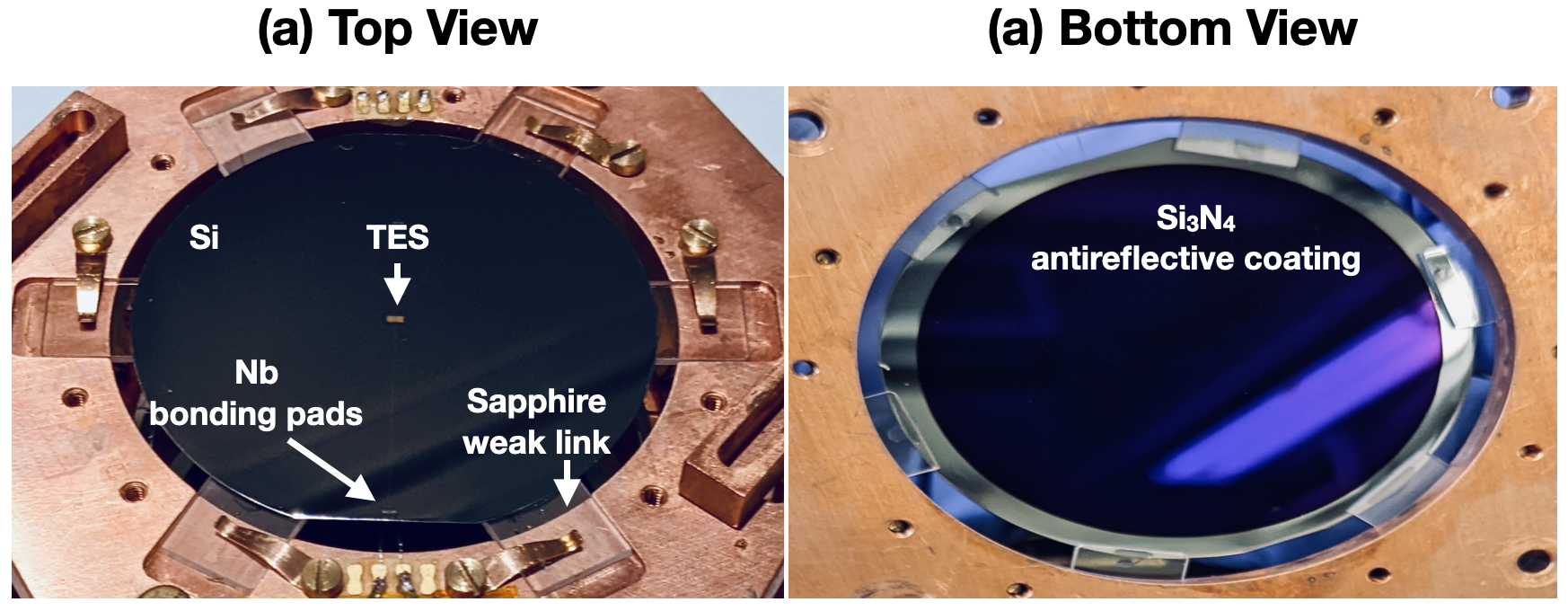}
\caption{\label{fig:detector} Low-temperature photon calorimeter. (a) Top view of Detector-I: The Ir-Pt bilayer sensor (TES) is located in the center of the Si wafer. The superconducting Nb traces connect the sensor to the bonding pads at the edge of the wafer. Single-crystal sapphire plates are used to provide a weak thermal link to the copper bath. (b) Bottom view of Detector-II where 68~nm Si$_{3}$N$_{4}$ has been deposited as an AR coating. Detector-II is identical to Detector-I except for the AR coating (see the text for details).}
\end{figure}

\begin{figure}[h]
\includegraphics[width=\linewidth]{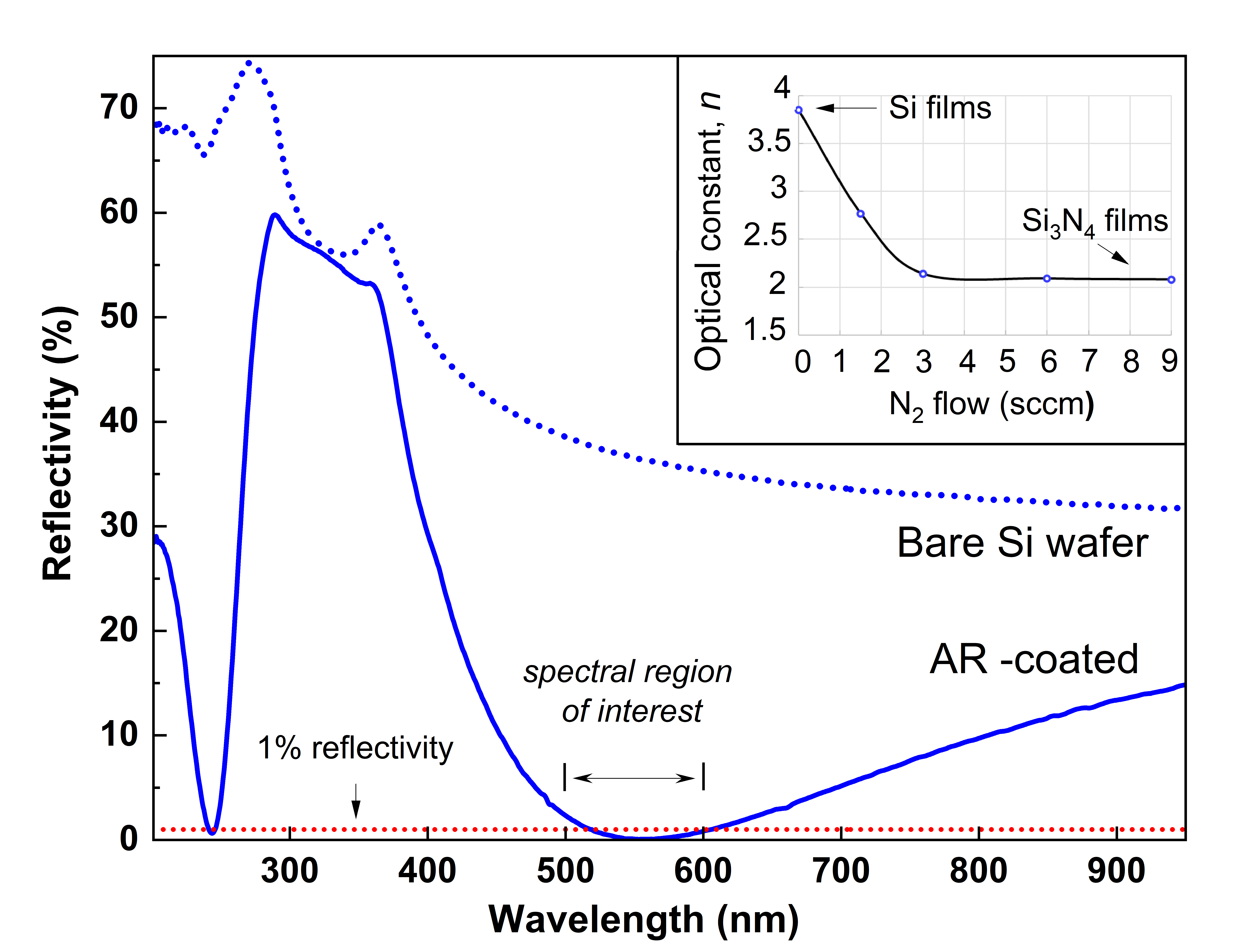}
\caption{\label{fig:reflectivity} Measured reflectivity of a Si wafer with 68~nm Si$_{3}$N$_{4}$ on top of it (solid blue line). The reflectivity in the spectral region of interest is below 1\% (dotted red line). For comparison, we also show the reflectivity of the bare Si wafer (dotted blue line). }
\end{figure}

We use high resistivity (>10000~$\Omega\cdot$cm) intrinsic silicon wafers (diameter = 50.8~mm and thickness = 280~$\mu$m) as calorimeters (Fig.~\ref{fig:detector}). Silicon can be produced with high radiopurity and has a relatively high Debye temperature $\Theta_{D}\sim$640~K~\cite{LandoltBornstein2002:sm_lbs_978-3-540-31356-4_478}, which ensures a marginal contribution to the total heat capacity of the calorimeter. We have fabricated several devices but selected two devices, Detector-I and Detector-II, for detailed characterization reported here. Both detectors are similar, with the difference being that Detector-II has a layer of Si$_{3}$N$_{4}$ anti-reflective (AR) coating deposited on the side of the wafer opposite the TES. P. Chen et al.~\cite{chen2018bridgman} have measured the luminescence spectra of LMO crystals down to a 20~K temperatures and found a broad peak around 551~nm with the luminescence intensity increasing with decreasing temperature. We chose Si$_{3}$N$_{4}$ as an anti-reflecting coating material due to its excellent optical properties, chemical stability, and natural compatibility with Si substrates. An added benefit is that thin films of this material are widely used by the microfabrication community and can be patterned via Reactive Ion Etching, which may be required in the future stages of the project. We sputtered the films reactively in a UHV-grade deposition chamber (base vacuum of ~1$\times$10$^{-8}$ Pa) using a DC power supply and a high-purity (99.999\%, undoped) Si target. We applied low power (20~W) RF bias for \textit{in situ} substrate cleaning, as a part of the pre-sputtering protocol, and during the deposition process. As-grown films are very smooth; we found that the average roughness measured by an Atomic Force Microscope was <1~nm. Furthermore, we confirmed excellent adhesion properties by inspecting the structural integrity of the samples after sonication in IPA and water for a prolonged time (30~min). We optimized the composition of thin films by varying the flow of N$_{2}$, while keeping the Ar flow, the deposition power, and the working pressure fixed at 10 sccm, 150~W and 3~mTorr, respectively. We measured the reflectance spectrum in the UV-VIS range using the Filmetrics F20 tool to determine the optical properties of the deposition films. As shown in the inset of Fig.~\ref{fig:reflectivity}, we can fine-tune the optical constant of the sputtered films in a wide range by adjusting the nitrogen flow. Based on these data, we have chosen to grow nitride films with a N$_{2}$ flow of 10~sccm, resulting in a refractive index n= $\sim$2.05 (at 632~nm), which is close to the values typical for stoichiometric Si$_{3}$N$_{4}$ composition. As shown in Fig.~\ref{fig:reflectivity}, $\sim$68~nm of thin material provides reflectivity of <1~\% in the entire range of interest (500--600~nm), a significant improvement compared to (35--40)\% for bare (uncoated) Si.    

\begin{figure}[h!]
\includegraphics[width=0.8\linewidth]{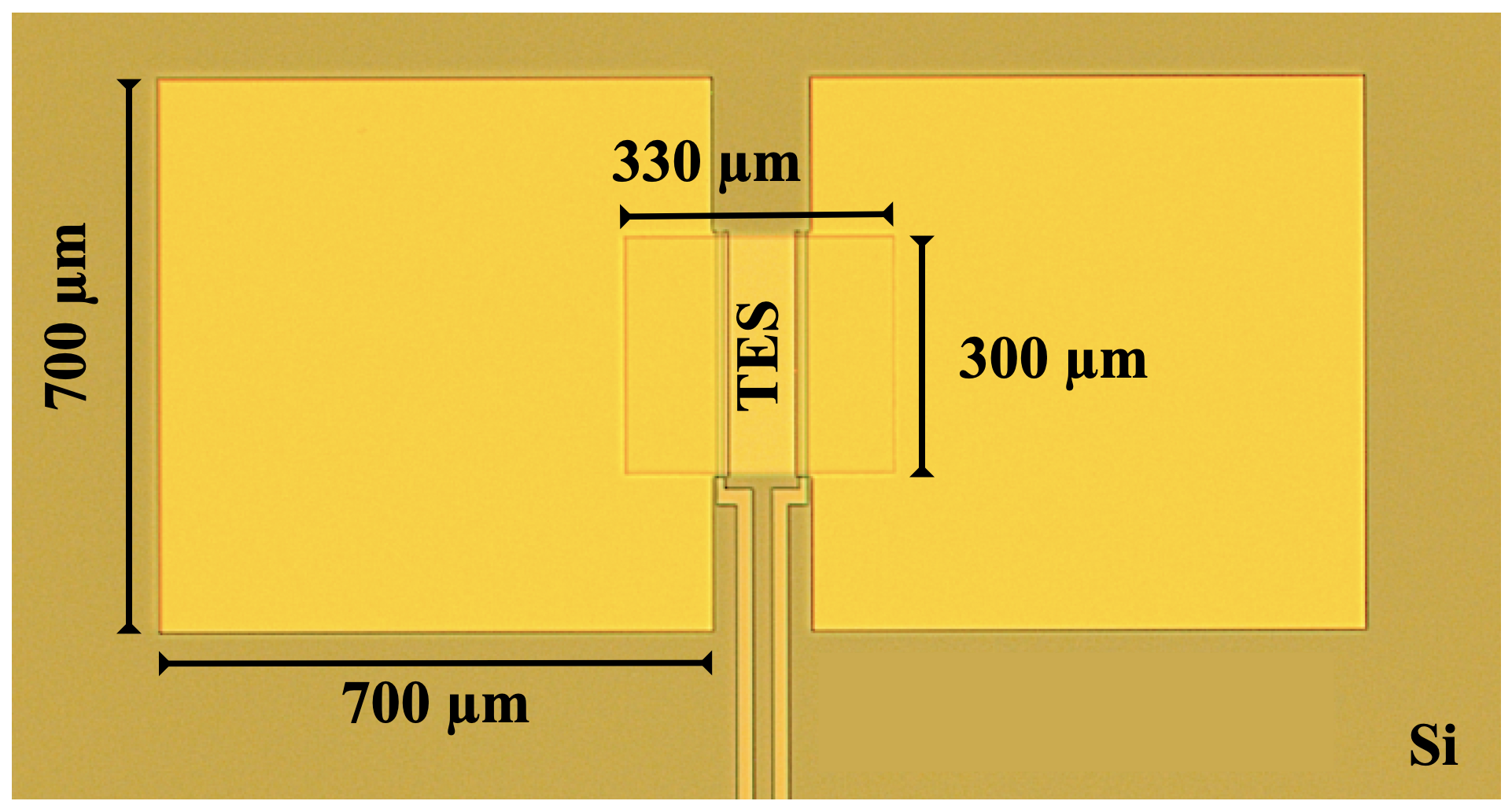}
\caption{\label{fig:schematic_tes} Schematic of the Ir(45~nm)-Pt(20~nm) bilayer sensor on the wafer. The Au pads overlap a portion of the TES and Si photoabsorber in order to improve the thermal coupling of the TES to the absorber.}
\end{figure}

We instrumented the Si absorber with an Ir-Pt bilayer TES close to the center of the Si wafer. The bilayer TES is a 45 nm Ir film deposited atop a 20 nm Pt film, both with an area of 330~$\mu$m $\times$ 300~$\mu$m. We then deposit Au pads (700~$\mu$m $\times$ 700~$\mu$m) on both sides of the bilayer to increase the phonon coupling between the sensor and the absorber.  The Au pads have a thickness of 200~nm and cover an area of 120~$\mu$m $\times$ 300~$\mu$m on each side of the bilayer. The choice of Au pad size was reached after studying its effect on the sensor-absorber coupling as described in Appendix~\ref{appedixA}. The remaining area of 90~$\mu$m $\times$ 300~$\mu$m was not covered with Au to obtain a normal resistance (R$_n$) between $\sim$(0.5 -- 1)~$\Omega$. The expected critical temperature of the sensor is $\sim$50~mK. Fig.~\ref{fig:schematic_tes} shows a schematic of the TES patterned on the Si absorber. 

We use superconducting Nb traces (200~nm thick and 20~$\mu$m wide) as electrical leads that run from the sensor edges and terminate as wire bonding pads near the edge of the wafer (Fig.~\ref{fig:detector}). The Nb leads are on top of the Au pads overlapping the sensor. We use 32~$\mu$m diameter Al-1\%Si wire bonds to connect the TES to the electrical circuit.  A thin gold strip (1000~$\mu$m $\times$ 100~$\mu$m $\times$ 150~nm) is deposited some distance away from the TES, for use as a Joule-heater. This heater has a measured residual resistance of 1~$\Omega$ below 1~K temperature. 

We use sapphire (Al$_2$O$_3$) crystals as the support (weak) link between the Si wafer and the Cu holder which is coupled to the thermal bath. Bachmann et al.~\cite{bachmann1972heat} have shown that 1/3 of the weak-link heat capacity contributes to the total heat capacity of the calorimeter. Therefore, we compute the additional heat capacity of the sapphire ($\Theta_{D}\sim$1042~K~\cite{doi:10.1063/1.322119}) to be <10\% of the Si absorber. Next, we glue the wafer to the sapphire plates using a tiny dot of UV-cured glue to restrict its horizontal plane movement. The sapphire, in turn, is held to the Cu holder using copper clamps, as shown in Fig.~\ref{fig:detector}. We expect the thermal boundary impedance between Cu and sapphire to be much larger than the bonded contact between Si and sapphire~\cite{doi:10.1063/1.1136802}. This ensures that heat does not leak out to the thermal bath before reaching the sensor.

\section{Measurements and Data Analysis}
We performed the measurements in an Oxford Instruments Triton 400 dilution refrigerator with a cryogen-free pulse tube cooler that reaches temperatures down to $\sim$10~mK. Each TES is operated with a near constant voltage bias to stabilize the device in the superconducting transition through negative electrothermal feedback~\cite{irwin1995quasiparticle}. To achieve a voltage biased state, we utilize shunt resistors that have a much lower resistance than the TES at its typical operating point ($R_{sh}$ $\sim$~20~m$\Omega$, $R_{TES}$ $\sim$ 50\%~$R_{n}$). We acquire the TES signal after amplifying it using a single-stage DC-Superconducting Quantum Interference Device (SQUID) array and readout electronics from Magnicon~\cite{magnicon}. The DC-SQUID and the shunt resistor are mounted in the Still stage of the dilution refrigerator, which operates at $\sim$ 700~mK. We use superconducting NbTi wires from the 700~mK stage to the 10~mK stage to bias the TESs. We monitor the temperature of the experimental stage using a Johnson-Noise thermometer~\cite{engert2012noise}. 

We measured a T$_c$ of $\sim$33~mK for both of our detectors and a $R_{n}$ of $\sim$550~m$\Omega$. The T$_c$'s are substantially lower than the expected value of $\sim$50~mK. This is not surprising because the TES is very weakly thermally coupled to the thermal bath. Therefore, any parasitic power, through RF/EMI or vibrations, can keep the wafer at a higher temperature than the measured bath temperature. As CUPID is designed to operate at ($\sim$~10--15~mK), we operate our detectors at a base temperature of 15~mK.

 We used two 200~$\mu$m core multimode fiber optic cables ($\lambda$ $\approx$ 400 -- 2200~nm), one for each detector, inside the cryostat to send light pulses from outside to calibrate the detector. The fiber ends face the opposite side of the wafer where the TES is deposited and have approximately normal angle of incidence.  We use an off-the-shelf commercial LED with a wavelength of $\sim$600~nm and an emission width (FWHM) of $\sim$12~nm. The LED optical fiber source is uncalibrated; however, it provides valuable information regarding timing and amplitude-dependent pulse shapes and can be used for energy calibration based on Poisson statistics of the photons impinging on the detectors. We do not have an infrared filter for the fibers inside the refrigerator. Although we have taken the utmost precaution to thermalize the fibers at different stages of the dilution refrigerator, some infrared radiation may leak into our detector through the fiber. The Au-heater is prone to injecting excess heat due to ambient RF/EMI interference and possible ground loops. This makes utilization of the fibers to excite the detector a much better option (see Appendix~\ref{appedixB}). 

\begin{figure}[ht]
\includegraphics[width=\linewidth]{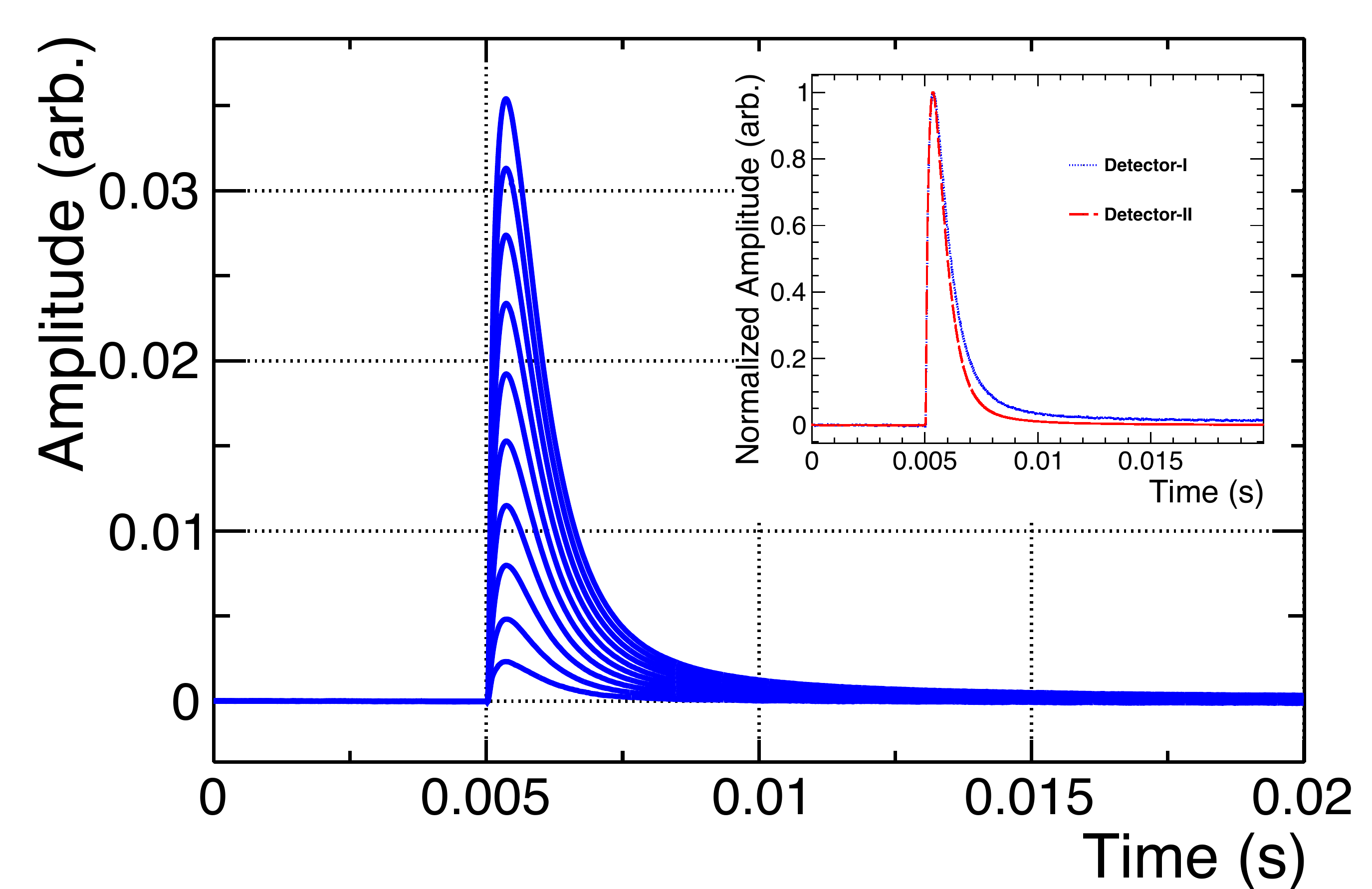}
\caption{\label{fig:avgPulses} Average pulse shapes for different LED excitations on Detector-I in the 20~ms event window. The inset shows the average pulse shapes for Detector-I and Detector-II for similar amplitudes. Detector-I pulses have typical risetimes of (175$\pm$10)~$\mu$s and decaytimes of (1000$\pm$40)~$\mu$s. Detector-II pulses have similar risetimes of (176$\pm$7)~$\mu$s but slightly faster decaytimes of (780$\pm$30)~$\mu$s. See the text for the definition of risetime and decaytime.}
\end{figure}

We generate a train of LED light pulses of varying amplitudes with a known separation time to test the detector response to the light signal. We choose the separation time between the generated LED pulses so that they are far enough in time to allow the detector to return to its quiescent state. We keep the voltage across the LED constant and vary the voltage pulse width across the LED to get a different number of photons for each pulse. As a result, the number of photons emitted from the LED has a linear dependence on its pulse width.  The width of the LED voltage is varied between 40 -- 400~ns. The pulse widths of excitation are much smaller than the typical risetime of the detectors (see Fig.~\ref{fig:avgPulses}), and, therefore, we treat all the LED excitations as instantaneous. In order to accurately identify these LED induced events, we acquire a synchronous signal from our LED voltage generator in order to provide a tag for these events in our data stream. This allows us to reject events from ambient background radioactivity (x-rays, muons, etc). 

The analog voltage of the SQUID output goes through a 10~kHz anti-aliasing filter before being acquired by an 18 bit analog-to-digital converter board (NI PXI-6289) with a sampling frequency of 150~kHz. We use a custom C++ DAQ software to continuously acquire data streams and save them to disk to process offline. Offline processing is performed using {\sc Octopus}~\cite{Octopus}, which is a modern {\sc C++} and {\sc ROOT}~\cite{brun1997root} based framework, to analyze continuous digitized waveforms. We use a simple threshold-over-rms method to identify and trigger on thermal pulses, and inject triggers at fixed intervals in order to produce a sample of noise events. {\sc Octopus} closely follows the implementation of frequency-based ``optimal filter'' (OF) outlined in Ref.~[\onlinecite{golwala2000exclusion}]. The OF maximizes the signal-to-noise ratio for the estimation of the pulse amplitude, under the assumption that the ideal pulse shape is known and the noise is stationary~\cite{szymkowiak1993signal,fowler2016optimal}. The ideal pulse template for the optimal filter is created by averaging a selection of LED pulses of a certain amplitude. This is done after rejecting events with a large baseline excursion, elevated noise levels, and requiring that only one pulse exists in a 15~ms window after the measured triggered time (Fig~\ref{fig:avgPulses}). On the other hand, the average noise power spectrum is created using the noise events. The pulse amplitudes of the LED are then corrected for the slow thermal drift of the detector by linearly regressing the known amplitude versus the baseline values (cf. Ref.~[\onlinecite{alduino2016analysis}] ). {\sc Octopus} also calculates basic pulse shape parameters, on raw and OF-filtered pulses, which are used for data processing and selection. We list a few of these parameters, which will be heavily referenced in the rest of the paper. 
\begin{itemize}
    \item \texttt{Risetime}: defined as the time it takes the leading edge of the pulse to go from 10\% to 90\% of the maximum amplitude. Typically $\sim$175~$\mu$s for both detectors.
    \item \texttt{Decaytime}: defined as the time it takes for the falling edge of the pulse to go from 90\% to 30\% of the maximum amplitude. Typical values of $\sim$1~ms and $\sim$800~$\mu$s for Detector-I and Detector-II, respectively.
    \item \texttt{OFChisquare}: statistic to measure the difference between the pulse and the template pulse after OF. Typical values of $\chi^2/d.o.f \sim 1$.
    \item \texttt{OFDelay}: time offset used to correct the pulse position relative to the trigger time in the event window. Depends on the pulse amplitude and timing jitter and has typical values of $\sim$10--20~$\mu$s. 
\end{itemize}

\section{Energy response and baseline resolution}
\begin{figure}[ht]
\includegraphics[width=\linewidth]{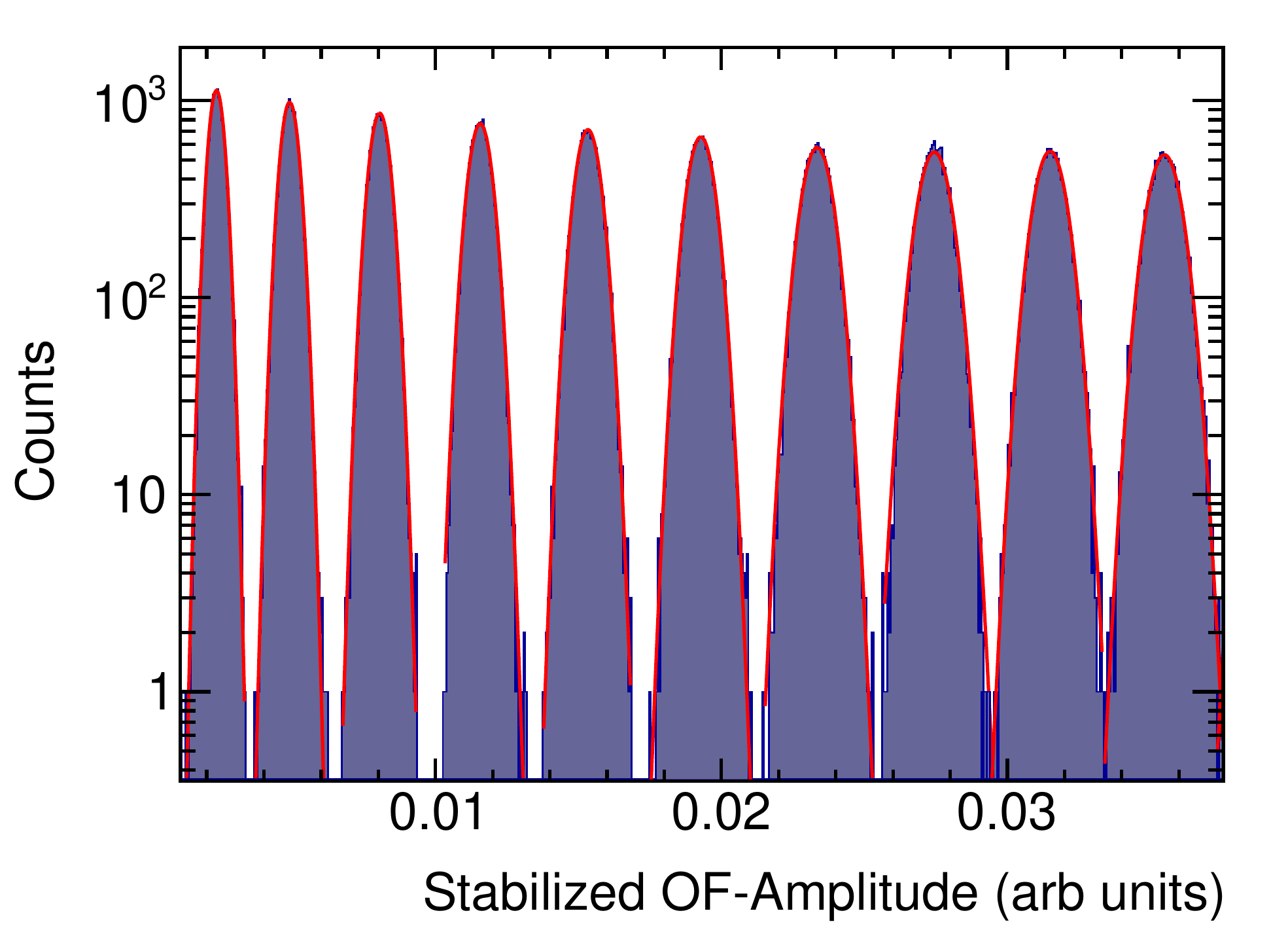}
\caption{\label{fig:amplitudeHisto} Amplitude histograms for LED excitations. We fit each peak with a Gaussian lineshape to obtain the centroid ($A_{i}$) and the width ($\sigma_{A_{i}}$).}
\end{figure}

We follow Cardani et al.~\cite{cardani2018ti, cardani2021final} to obtain an energy calibration based on the Poisson statistics of the LED photons absorbed in our detectors. Each LED trigger results in a certain number of photons ($N_{i}$) that deposit all of their energy in the Si absorber. This gives rise to a Gaussian distribution with a peak at the mean amplitude of $A_{i}$ and a width of $\sigma_{A_{i}}$ (Fig.~\ref{fig:amplitudeHisto}). Assuming a baseline resolution of $\sigma_{0}$, we can write 
\begin{equation}
   \sigma_{A_{i}}^2= \sigma_{0}^2 + \left( R \cdot \sqrt{N_{i}} \right)^2,
   \label{eqn:sigmaA1}
 \end{equation}
 where $R$ is the responsivity ({dA}/{dN}) of the detector per photon. Since $A_{i} = R*N_{i}$, Eq.~(\ref{eqn:sigmaA1}) reduces to 
 \begin{equation}
   \sigma_{A_{i}}^2= \sigma_{0}^2 + R \cdot {A_{i}}.
   \label{eqn:sigmaA2}
 \end{equation}

We obtain the responsivity $R$ by fitting $\sigma_{A}$ versus $A$ with the functional form shown in Eq.~(\ref{eqn:sigmaA2}). We then calculate the total number of photons $N_{i}$ for each $A_{i}$. Since the mean energy of each photon is known (2.1~eV for $\lambda$=600~nm), we get an absolute calibration for each amplitude by multiplying the corresponding $N_{i}$ by the photon energy. We treat the baseline noise as independent of amplitude and assume that the Poisson fluctuation of the number of photons dominates the stochastic term (second term in Eq.~(\ref{eqn:sigmaA1})). In Appendix~\ref{appedixB}, we show that these are true for our case.

\begin{figure}[ht]
\includegraphics[width=\linewidth]{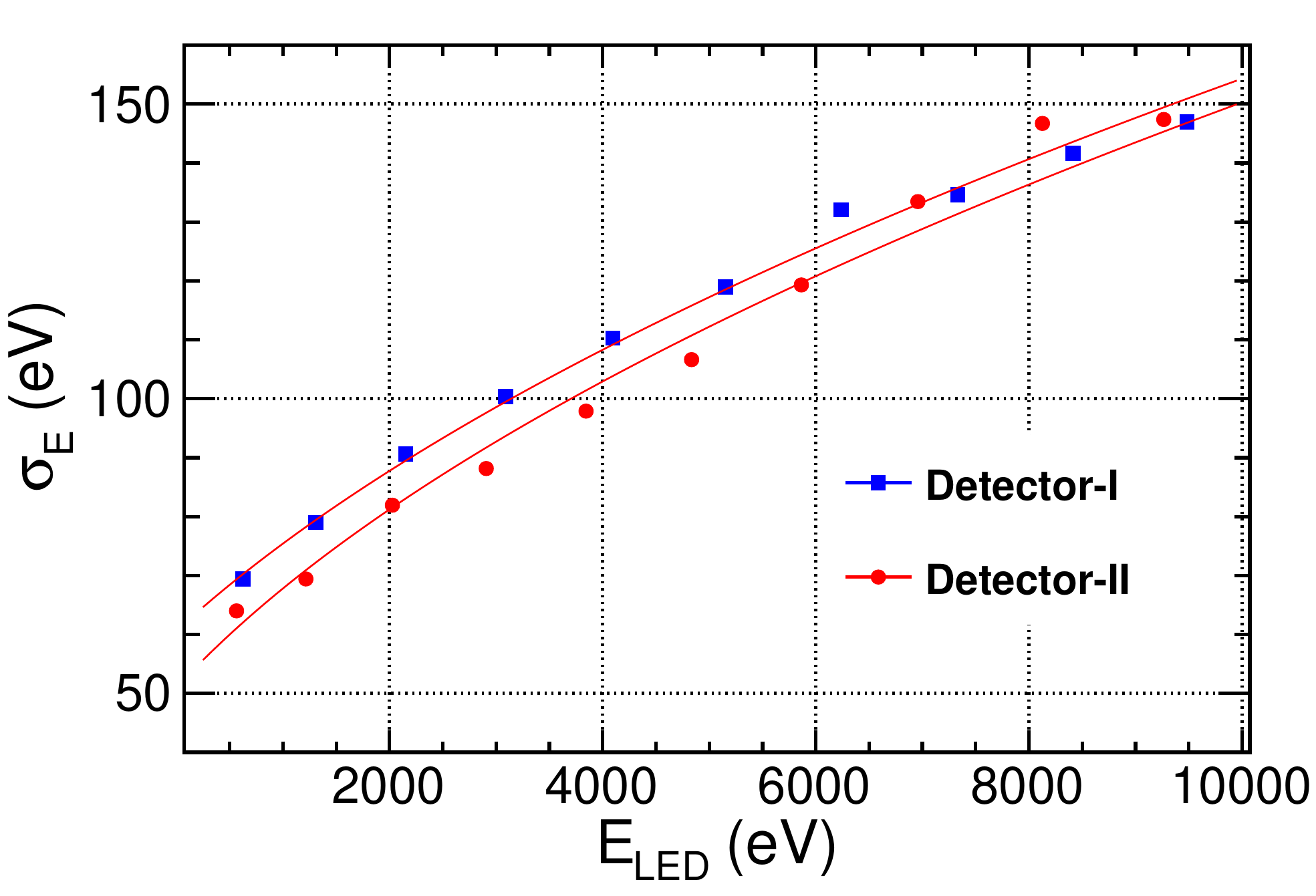}
\caption{\label{fig:energyresolution} $\sigma_{E}$ vs $E$ from LED excitation of Detector-I (without AR coating) and Detector-II (with AR coating) at 15~mK. The fitted function corresponds to Eq.~(\ref{eqn:sigmaE}) where $\sigma_{s}$ is assumed to be dominated by event-to-event fluctuations in the LED signal corresponding to Poisson statistics (see the text for details). The error bars are within the data markers, and we report the fit parameters in Table~\ref{tab:table1}. The deviation between each data point and the fitted curves can be explained by the presence of telegraphic noise, which is not accounted for by Eq.~(\ref{eqn:sigmaE}).}
\end{figure}

\begin{table}[ht]
\caption{\label{tab:table1}Fit results from the fit of data to Eq.~(\ref{eqn:sigmaE}) for LED excitations. $\sigma_{n}$ and $\sigma_{s}$ are moderately anticorrelated.}
\begin{ruledtabular}
\begin{tabular}{cccc}
 & {$\sigma_{n}$~(eV)}  & {$\sigma_{s}$~($\sqrt\text{eV}$)}  & {Correlation ($\sigma_{n}$,$\sigma_{s}$)} \\
 \hline
Det-I & 61$\pm$0.5 & 1.42$\pm$0.01  &  -0.73 \\
Det-II & 51$\pm$0.5 & 1.41$\pm$0.01  & -0.72 \\ 
\end{tabular}
\end{ruledtabular}
\end{table}

Once calibrated, $\sigma_{E}$ vs. $E$ is fitted again to 
 \begin{equation}
   \left( \frac{\sigma_E}{E} \right) ^2 =  \left (\frac{\sigma_{s}}{\sqrt{E}} \right)^2 +\left (\frac{\sigma_{n}}{E} \right)^2,
   \label{eqn:sigmaE}
 \end{equation}
where $\sigma_{n}$ is the baseline noise of the detector (independent of the amplitude) and $\sigma_{s}$ arises due to the stochastic nature of the primary excitation signals. 

We find that Detector-I has an $R_{n} \approx$ 480~m$\Omega$ and Detector-II has an $R_{n} \approx$ 650~m$\Omega$. We operate Detector-I at $\sim$0.35~$R_{n}$ and Detector-II at $\sim$0.50~$R_{n}$. We could not operate both detectors under similar bias conditions because we observed small discontinuities and kinks for many bias points throughout the transition. Similar observations of telegraph noise at certain bias voltages in TES and instabilities have been previously reported in Refs.~[\onlinecite{kotsubo2012observation,croce2012ultra,hatakeyama2014gamma}]. Bennet et al.~\cite{bennett2014phase} show that discrete changes in the number of phase-slip lines could be a possible explanation for the observed switching between discrete current states in localized regions of bias. The complete characterization of the transition, noise sources, and their mechanisms is currently under active study and will be discussed in a separate paper. For the work shown here, it was possible to bias the TESs in a narrow transition region, where the current-voltage characteristics were smooth for small current excursions.

 We can estimate the particle discrimination capabilities for $\alpha$/$\beta$ using the calibration curve shown in Fig.~\ref{fig:energyresolution}. CUPID needs a rejection factor >99.9\% to reduce the $\alpha$ background below $10^{-4}$~counts/(kg/keV/yr)\cite{poda2017low,beeman2012next}. We can quantitatively show that our detectors satisfy the CUPID requirements by using a classification metric for $\alpha$ and $\beta / \gamma$ distributions. We use the Bhattacharyya distance~\cite{bhattacharyya1943measure}, which is common for feature selection and is known to provide lower and upper bounds of the classification error for normal distributions. The Bhattacharya distance between two normal distributions can be written as 

 \begin{multline}
   d_{B} = \frac{1}{4} \cdot \frac{(\mu_{\beta/\gamma }(E)-\mu_{\alpha}(E))^2}{\sigma_{\beta / \gamma}^2(E) + \sigma_{alpha}^2(E)} \\ + \frac{1}{2}\cdot {\text {ln}} \left(\frac{\sigma_{\beta / \gamma}^2(E) + \sigma_{alpha}^2(E)}{2\cdot\sigma_{\beta / \gamma}(E)\cdot\sigma_{alpha}(E)}\right)
   \label{eqn:bhattacharya}
\end{multline}
 
 where $\mu$ and $\sigma$ are the average value and RMS of the $\beta / \gamma$ and $\alpha$ light distributions, respectively. Both $\mu$ and $\sigma$ depend on the energy ($E$) of the incident particle in the LMO. Assuming {\textit {a priori}} equal probabilities for both distributions, the upper bound for the Bayes classification error is given by~\cite{lee2000bayes,fukunaga2013introduction,kailath1967divergence}
 
  \begin{equation}
   \varepsilon \leq \frac{1}{2} \cdot {\text {e}}^{-d_B}
   \label{eqn:classierror}
 \end{equation}

At $\sim$3~MeV we expect an average light energy of 1~keV for $\beta/\gamma$ particles. The quenching factor for $\alpha$ scintillation in LMO~\cite{poda2021scintillation,armengaud2020cupid} is 0.2 and therefore an average of 200~eV is expected on the light detectors from $\alpha$'s. Using the respective $\sigma_E$'s (Fig~\ref{fig:energyresolution}) we obtain $\varepsilon \leq$~4.5$\times$10$^{-8}$ and $\varepsilon \leq$~3.4$\times$10$^{-10}$ for Detector-I and Detector-II, respectively, which demonstrates excellent discrimination capabilities. The exponential dependence in Eq.(\ref{eqn:classierror}) shows why having a small baseline resolution and a higher light yield is necessary; the classification error bound quickly increases to $\varepsilon \leq$~3.9$\times$10$^{-4}$ if we increase the baseline noise ($\sigma_{n}$) to 100~eV. Note that some of the previous publications~\cite{cupid_cdr,poda2017low,armengaud2020cupid} define the discrimination power (DP) using only the first term of Eq.~(\ref{eqn:bhattacharya}) with 
\begin{equation}
  {\text {DP}} = \frac{\left|\mu_{\beta/\gamma }(E)-\mu_{\alpha}(E)\right|}{\sqrt{\sigma_{\beta / \gamma}^2(E) + \sigma_{alpha}^2(E)}}
\end{equation}
and quote DP($E$=$Q_{\beta\beta}$)>3.1 as a requirement to reach an alpha-induced background rate of <~$1\times10^{-4}$~counts/(kg/keV/yr). For comparison, we cite a DP of 8.1 and 9.2 for Detector-I and Detector-II, respectively.

\begin{figure}[ht]
\includegraphics[width=\linewidth]{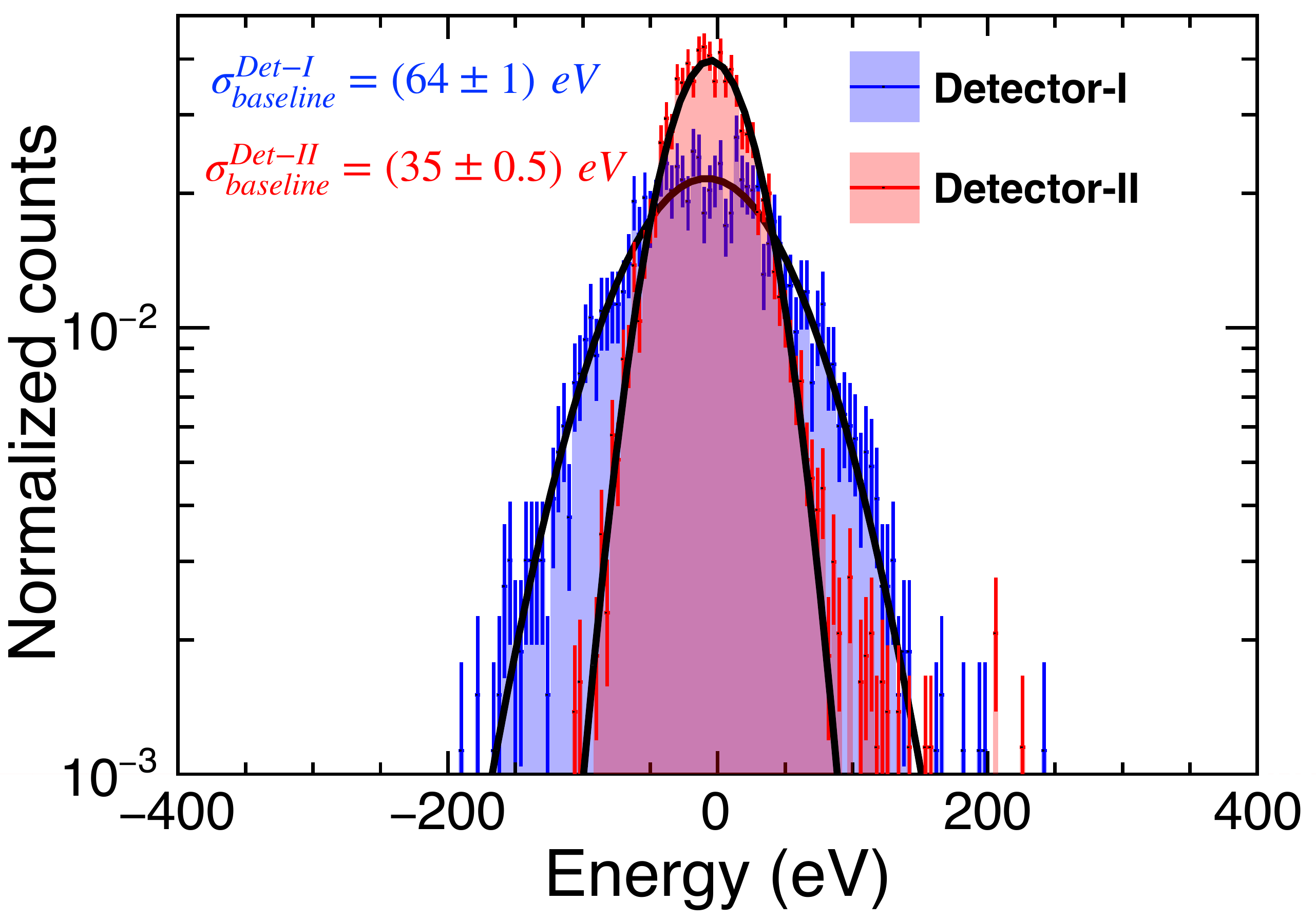}
\caption{\label{fig:baselineResolution} Energy distribution of noise triggers using LED calibration. We measured a baseline resolution of (64$\pm$1)~eV and (35$\pm$0.5)~eV for Detector-I and Detector-II, respectively. The noise distributions are fit to Gaussian lineshapes to obtain the respective widths ($\sigma_{baseline}$).}
\end{figure}

 We also calibrate the energy of the triggered noise events to obtain a baseline noise resolution. We obtain a baseline noise resolution of (64$\pm$)1~eV and (35$\pm$0.5)~eV at 15~mK for Detector-I and Detector-II, respectively (see Fig.~\ref{fig:baselineResolution}). The baseline resolution for Detector-I is consistent with $\sigma_{n}$ obtained from the calibration fit (Table~\ref{tab:table1}). However, Detector-II has a smaller baseline resolution than its $\sigma_{n}$. The smaller quiescent-state noise in Detector-II points to bias-dependent noise in its TES sensor. The difference in baseline energy resolution ($\sigma_{n}$) between Detector-I and Detector-II is due to the difference in their operating point and the higher noise in Detector-II rather than being an effect of the AR coating. We have verified that the AR coating does not adversely affect the energy resolution. The fit results are presented in Table~\ref{tab:table1}. Both detectors meet the CUPID criterion for $\sigma_{n}$ < 100~eV .

\section{Timing resolution and pileup discrimination}
\begin{figure}[ht]
\includegraphics[width=\linewidth]{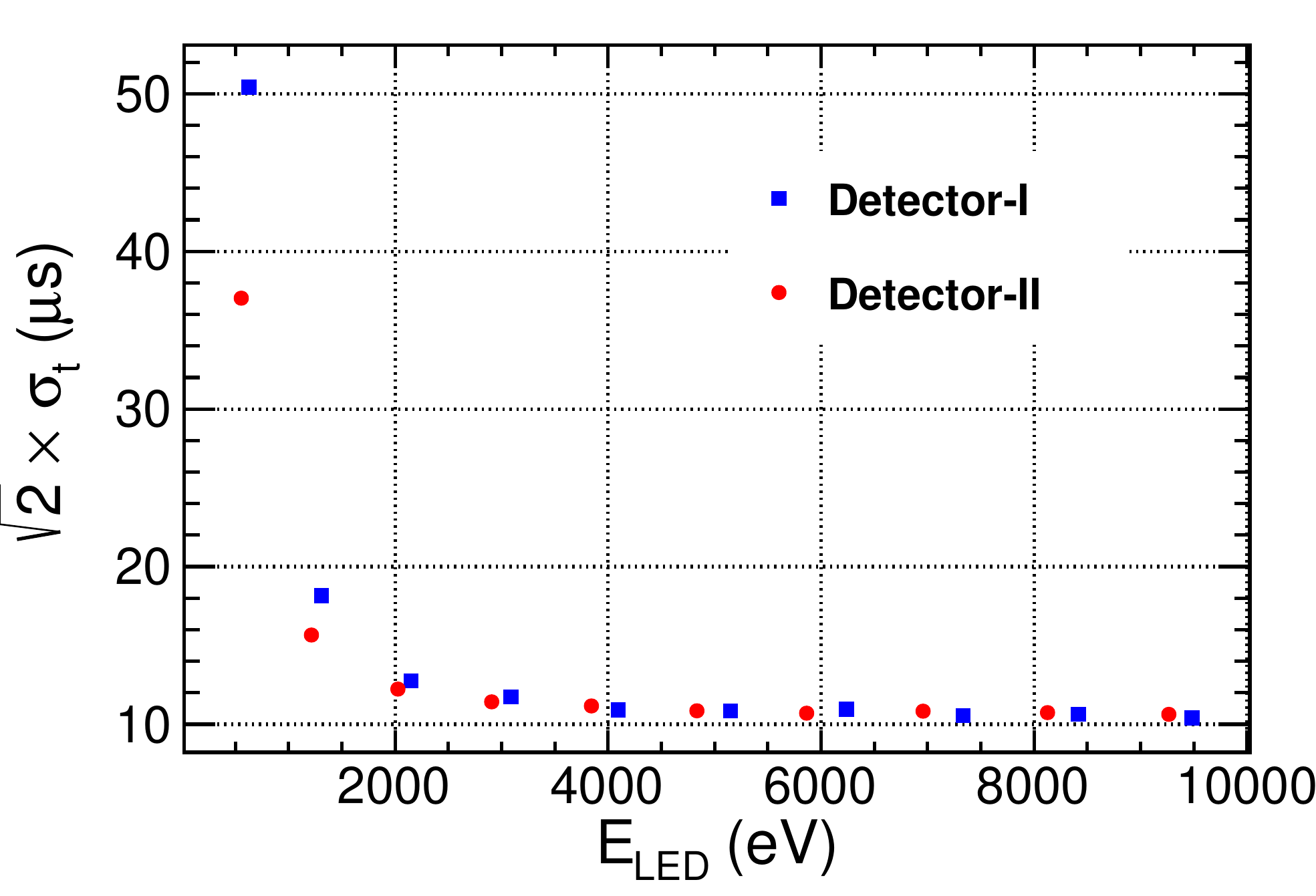}
\caption{\label{fig:timingresolution} Dependence of timing jitter $\sigma_{t}$ as a function of energy. The errors are within the marker size. For pulses with lower energy, where the SNR is poor, fluctuations in signal amplitude that cross the trigger threshold translate into additional timing fluctuations. The jitter value flattens out for high energy pulses because of the sampling rate (150~kS/s).}
\end{figure}

In timing measurements, the pulse rise time is important because the timing jitter depends on the slope-to-noise ratio rather than the signal-to-noise ratio alone. The timing jitter ($\sigma_t$) distribution can be written as 
 \begin{equation}
   {\sigma_t} =  \frac{\sigma_{n}}{({dS}/{dt})_{S_{T}}} 
   \label{eqn:sigmaT}
 \end{equation}
 where $\sigma_{n}$ is the rms noise and the derivative of the signal (${dS}/{dt}$) is calculated at the position of the trigger~\cite{spieler2005semiconductor}. To measure timing jitter, we use LED pulses to provide a start and stop signal for a known time interval and measure the spread in the distribution of $\Delta T = t_{start} - t_{stop}$. Fig.~\ref{fig:timingresolution} shows the measured timing jitter as a function of the pulse amplitudes for both detectors. The timing jitter is high for low-energy pulses because noise affects the time at which a trigger is issued. Nevertheless, we have measured a $\sigma_t$ < 20 $\mu s$  for all energies >~1~keV, which converges to $\sim$10~$\mu s$ above 3~keV. In the future, the timing jitter for low-energy pulses can be further reduced using an optimal triggering algorithm~\cite{di2011lowering}. 
 
Timing jitter of $\sigma_t$ < 20 $\mu s$ is enough for the CUPID baseline experiment, where we need to discriminate two $2\nu\beta\beta$ pileup events that occur as close as $\sim$170~$\mu s$ from each other. To show the discrimination power of the detector, we did a toy experiment in which we produced pileup pulses by generating two LED pulses temporally close to each other and comparing them to single-pulse signal events. We build a signal-like event sample by carefully selecting single LED pulse events of $\sim$1~keV energy. We characterize each event by five pulse-shape parameters and represent each event with a point in this five-dimensional space. The five pulse shape parameters are \texttt{Risetime}, \texttt{Decaytime}, \texttt{OFChisq}, \texttt{OFDelay}, and the sum of the integral under the rising edge and the falling edge of the pulse normalized to the total integral of the pulse from 10\% of amplitude on the rising edge to 30\% of amplitude on the falling tail. The variables are dependent on the pulse shape, which we expect to be modified in the presence of pileup events.  Identification of a pile-up event is then reduced to the problem of outlier detection in a signal-like space. We use the Mahalanobis distance ($d_{M}$) ~\cite{mahalonobis1936generalized}, a measure of the distance of a multidimensional point ($\vec x$) from a multivariate normal distribution that has a mean $\vec \mu$ and the covariance matrix ${\Sigma}$. We then define the outlier detection criterion by choosing a distance $k$ that satisfies 

\begin{equation}
d_M = \sqrt{{(\vec x-\vec \mu)^{T}}{\Sigma^{-1}} {(\vec x-\vec\mu)}}~ > ~k,
\label{eqn:MDist}
\end{equation}

where $k$ is chosen according to the number of variables used to build the distance. The distance inherently considers the correlations between the parameters and is always positive. We use Detector-I for the pileup experiment since it has a larger baseline noise and higher timing jitter. We note that if Detector-I is able to meet the CUPID criterion for timing, then Detector-II should as well.

\begin{figure}[ht]
\includegraphics[width=\linewidth]{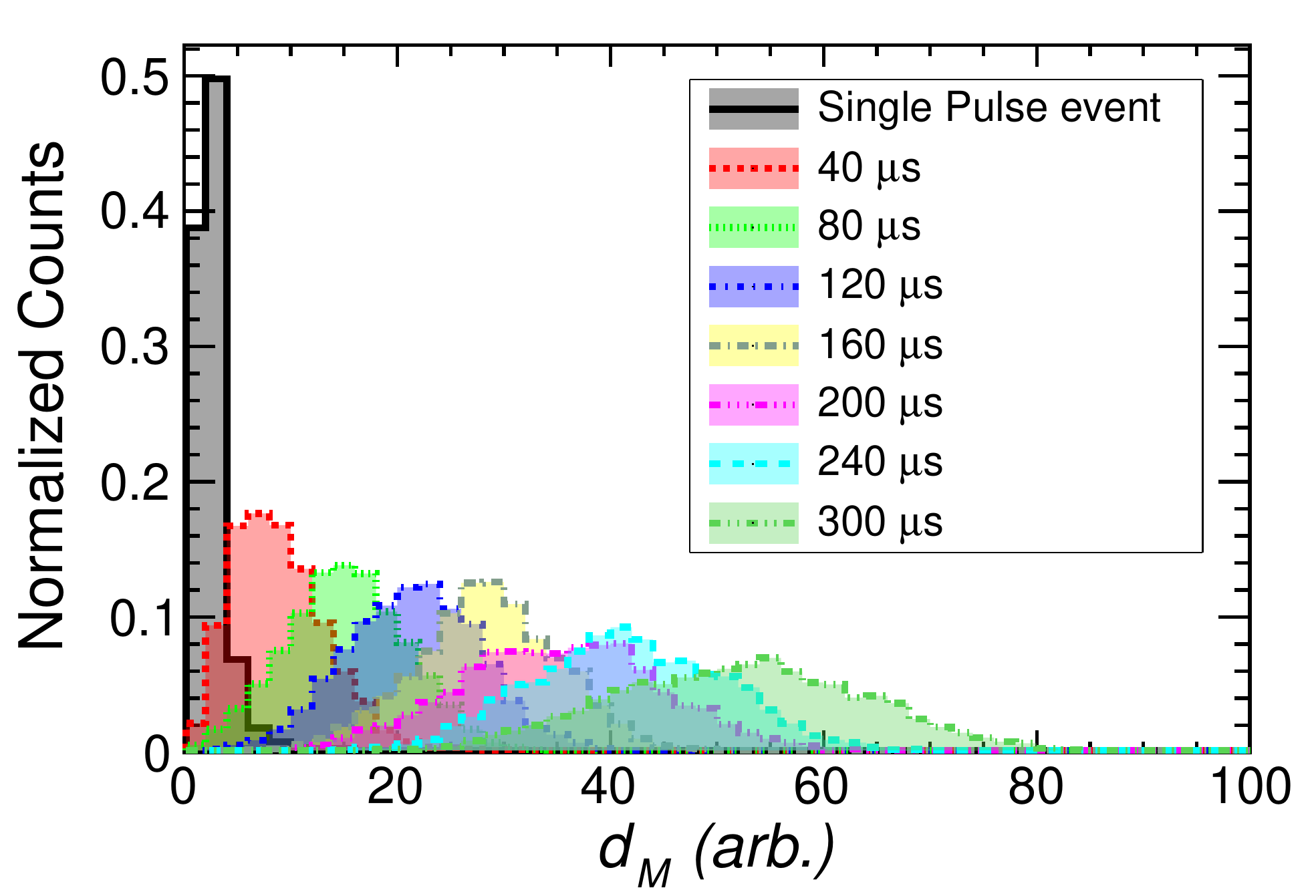}
\caption{\label{fig:mdistDistribution} Mahalanobis distance ($d_{M}$) distributions for single-pulse events and double-pulse events (similar to pileup) as a function of the time separation between double pulses.}
\end{figure}

\begin{figure}[ht]
\includegraphics[width=\linewidth]{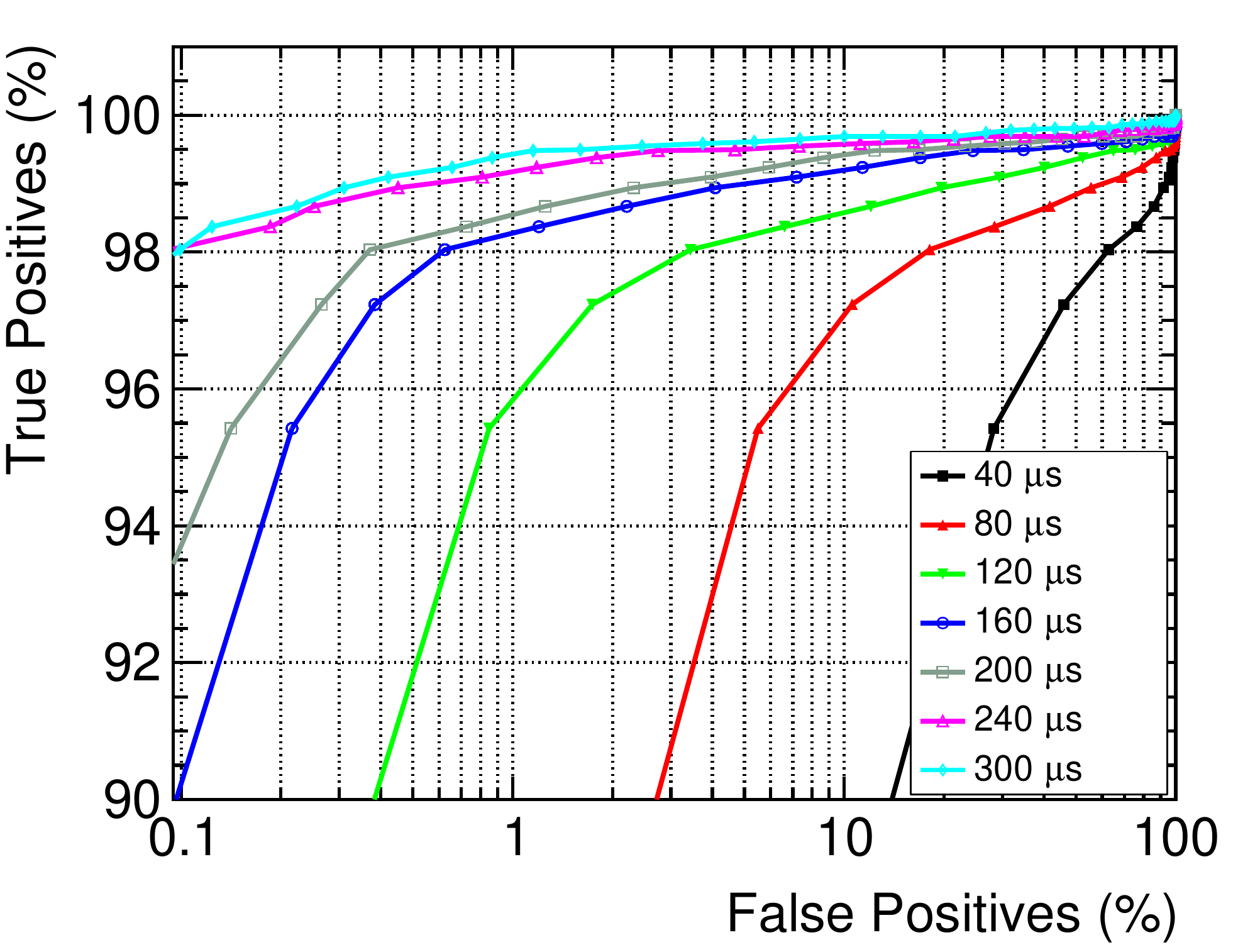}
\caption{\label{fig:mdistROC} ROC curves as a function of the time separation between double pulses. The ROC curves are obtained using the selection criteria of Eq.~(\ref{eqn:MDist}) where $k$ corresponds to the locus of a point that is a certain distance from the false positive axis. For a large $k$ we get an increase in the true positive rate, but also admit a higher number of false positive events. We obtain a precision of >99\% for pileup events that are more than 160~$\mu$s apart. }
\end{figure}

The distance distribution for single-pulse events and pile-up-like double-pulse events is shown in Fig.~\ref{fig:mdistDistribution}, while the Receiver Operating Characteristic (ROC) curves as a function of temporal separation between pileup-like double pulses are shown in Fig.~\ref{fig:mdistROC}. Each point in the ROC curve corresponds to a predetermined distance cut $k$ which corresponds to a certain constant distance from the False Positive axis shown in Fig.~\ref{fig:mdistROC}. An increase in $k$ increases the acceptance of single pulse events from its distribution (see Fig.~\ref{fig:mdistDistribution}) but at the cost of letting in more pile-up like events. The precision of our $d_{M}$ based selection criteria is defined as

\begin{equation}
    Precision~=~\frac{True Positives}{True Positives + False Positives}.
\label{eqn:precision}
\end{equation}

\begin{figure}[ht]
\includegraphics[width=\linewidth]{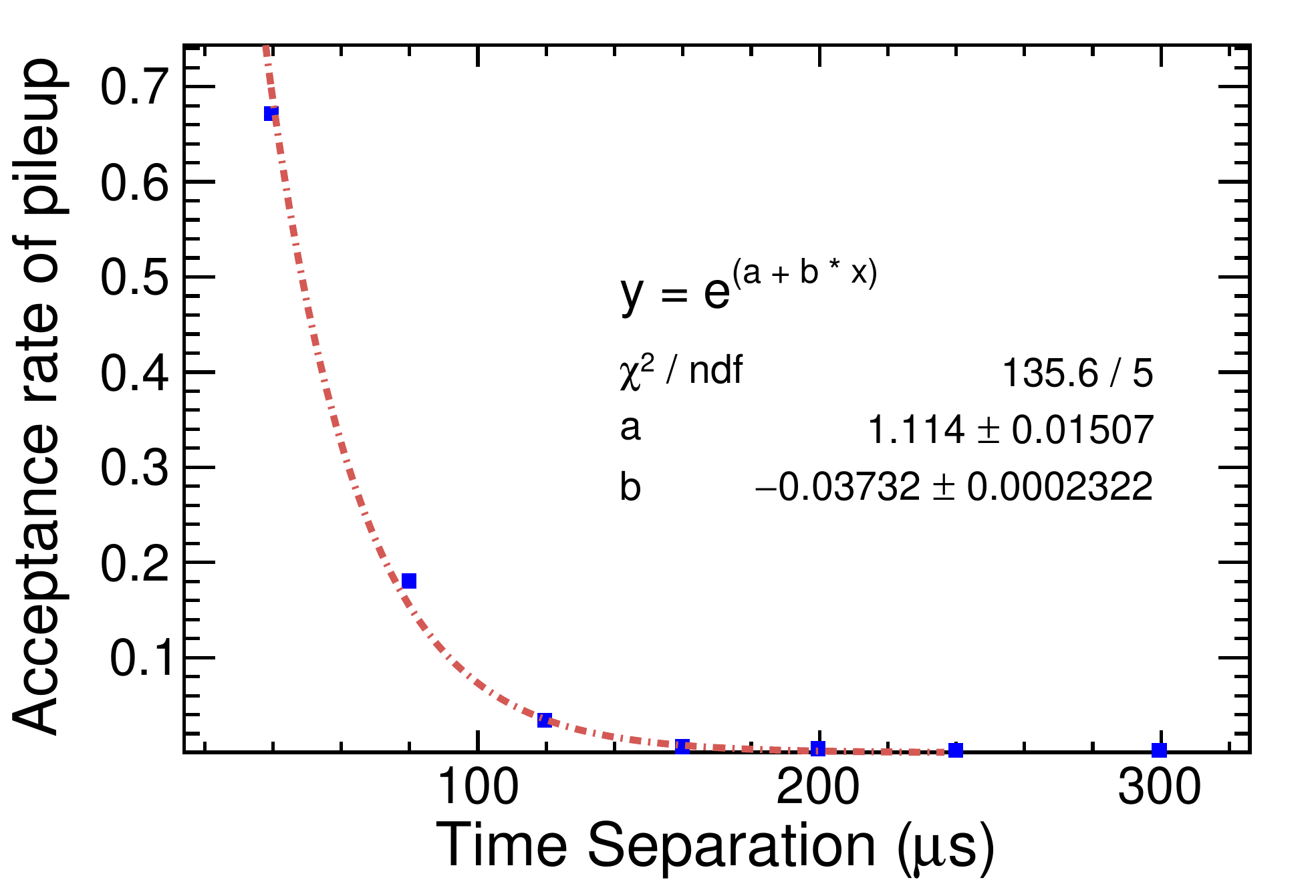}
\caption{\label{fig:acceptanceRate} Acceptance efficiency of pileup pulses (false positives) as a function of the time separation between them. Data points correspond to 98\% signal efficiency for single pulse events (see Fig.~\ref{fig:mdistROC}). We use an exponential for parametric fitting and calculate the area under the curve for the net acceptance efficiency from 40~$\mu$s to 1~ms. In the absence of measurements, we assume an acceptance efficiency of 100\% below 40~$\mu$s. }
\end{figure}

 Due to the fast rise time of our detectors, we can effectively discriminate pileup events within $\sim$120~$\mu s$ with a signal efficiency of 96\% and accept only 1\% of pileup events (precision $\sim$99\%). For pileup events that occur with a separation of at least $\sim$160~$\mu s$, we obtain a precision of > 99\%. We benchmark our pileup discrimination capabilities by calculating an expected background rate using the data presented in Fig.~\ref{fig:mdistROC}. Chernyak et al.~\cite{chernyak2012random} show that a background of $3.3\times10^{-4}$ counts/(kg/keV/yr) is expected for unresolved pileups below 1~ms in the region of interest of $^{100}$Mo. We set the acceptance efficiency for a single pulse event at 98\% and plot the corresponding acceptance efficiency for pileup events as a function of the separation time (Fig.~\ref{fig:acceptanceRate}). We use an exponential function to perform a parametric fit of the data. The total acceptance rate between [0,1~ms] is the area under the fitted curve. Since we do not have any measurements below 40~$\mu$s, we take a conservative approach and assume an acceptance rate of 100\% below 40~$\mu$s. Adding it to the area under the curve from 40~$\mu$s to 1~ms, we get a total acceptance rate of 5.8\% and hence a pileup background rate of $1.9\times10^{-5}$ counts/(kg/keV/yr). This is about three times lower than the CUPID target of $5\times10^{-5}$~counts/(kg/keV/yr). We caution the reader that this exercise is merely a benchmark for the light detectors and not the actual achievable background rate since we have performed our toy experiment with equal amplitude pulses of 1~keV. In practice, pileup events can have different amplitudes. Nevertheless, we have shown that our current generation of detector is adequate to reach CUPID targets. We are working towards improving our pileup discrimination capabilities by using machine learning techniques that can achieve higher precision.
 
\section{Conclusion}
We have demonstrated that a novel bilayer transition edge sensor-based light detector is a promising technology for CUPID. The detectors can be operated down to 10~mK and meet the energy resolution requirement of $\sigma_{n}$< 100~eV. The large surface area of the detector is crucial for detecting a low light yield from a scintillating LMO crystal. Furthermore, we have shown that adding a thin amorphous Si$_3$N$_4$  antireflective coating does not degrade detector performance. In this work, we have also shown how we can tune the absorber-sensor coupling and engineer the thermal conductances of the detector. More importantly, we have achieved an excellent timing resolution and have demonstrated that the current generation of detectors will suffice to meet the CUPID requirements of an effective timing resolution of <170~$\mu s$. For our next steps, we plan to produce more devices to understand the microscopic physics of the TES and energy thermalization and the source of excess noise in our TES. Work is underway to implement a frequency division multiplexing readout scheme and to demonstrate its feasibility for the CUPID experiment.

\begin{acknowledgments}
This work was supported by the US Department of Energy (DOE), Office of Science under Contract No. DE-AC02-05CH11231 and DE-AC02-06CH11357, and by the DOE Office of Science, Office of Nuclear Physics under Contract No. DE-FG02-00ER41138. Superconducting thin films synthesis was supported by the DOE Office of Science, Office of Basic Energy Sciences, Materials Sciences and Engineering Division. Use of the Center for Nanoscale Materials, an Office of Science user facility, was supported by the DOE Office of Science, Office of Basic Energy Sciences under Contract No. DE-AC02-06CH11357.
\end{acknowledgments}

\section*{Data Availability Statement}
The data that support the findings of
this study are available from the
corresponding author upon reasonable
request.

\appendix
\section{Effect of Au pad on the thermal conductance of TESs}\label{appedixA}
\begin{figure}[ht]
\includegraphics[width=0.8\linewidth]{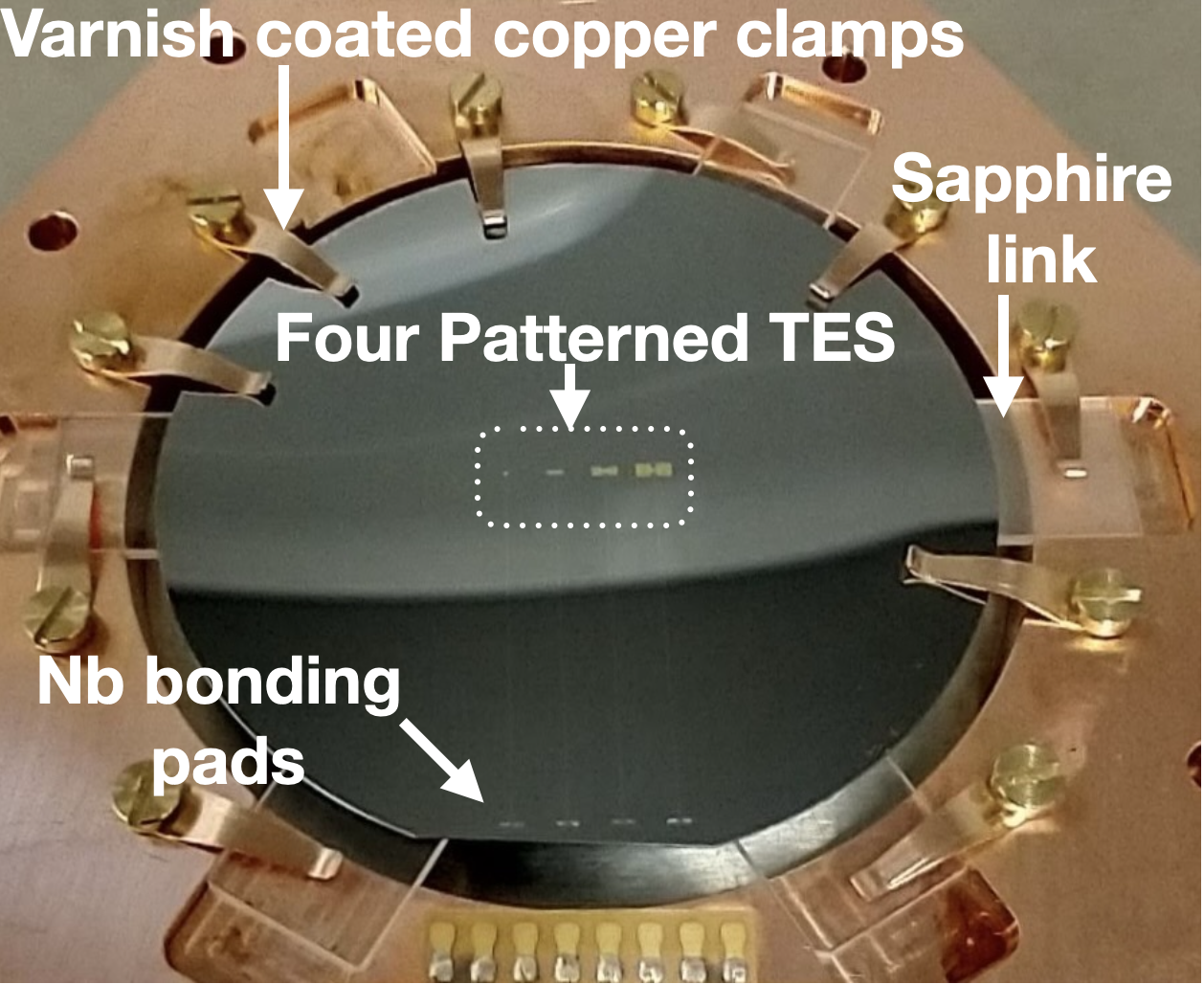}
\caption{\label{fig:detector4sensor} Detector with four TESs with different areas of Au that overlap the sensors and the Si absorber. Additional copper clamps were added to the Si wafer to increase the thermal coupling between the wafer and the copper bath and make $T_{absorber} \approx T_{bath}$. }
\end{figure}

The addition of Au pads was a crucial step in the construction of our device. Our initial tests of a wafer with a bare Ir-Pt TES had a small signal size with a poor signal-to-noise ratio. This was because of the weak thermal coupling between the sensor and the absorber, which was limited by the e-ph coupling of the bilayer TES. In a separate experiment, we studied how thermal coupling can be improved by using overlapping Au pads between the TES and Si absorber. 
We instrumented a Si absorber with four bilayer TES sensors close to the center of the Si wafer (Fig.\ref{fig:detector4sensor}). Each TES has a 100~nm thick Ir film deposited on top of a 50~nm thick Pt film, each with an area of 300~$\mu$m $\times$ 300~$\mu$m. We then deposit Au pads on both sides of the TES to increase the phonon coupling between the sensor and the absorber. The Au pads have a thickness of 200~nm. We deposit Au of different surface areas for each sensor (see Table~\ref{tab:table2}) to systematically study the effect of Au on the coupling between the absorber and the sensor. 

The typical procedure to measure the thermal conductance of a TES, $G(T_{tes})$, involves measuring Joule power heating as a function of bath temperature ($T_b$). We used the current-voltage (I-V) curves to obtain the Joule power, for a particular resistance, at each bath temperature. The power curve thus obtained can be fitted to the following:
\begin{equation}
   P_e(T)~=~k\cdot(T_{e,tes}^{n} - T_{b}^{n})
   \label{eqn:PvsT}
\end{equation}
where $P_e(T)$ and $T_{e,tes}$ are the power dissipated in the TES electron system and its temperature, respectively.

\begin{table}[ht]
\caption{ \label{tab:table2}Size of Au pad and $T_c$ for each of the four sensors shown in Fig.~\ref{fig:detector4sensor}. Each sensor is flanked by Au film of the mentioned size on either side. See text for a detailed discussion on thermal conductance between TES and absorber, $G_{ea}$. In the absence of an Au layer, Sensor-A is susceptible to parasitic power, which explains the lower measured $T_c$. The Thevenin equivalent circuit parameters for Sensor-C TES readout were not precisely known and were assumed to be similar to the other circuits. This may explain the abnormally high $R_{n}$ measured for Sensor-C.} 
\begin{ruledtabular}
\begin{tabular}{ccccc}
      \textbf{Sensor} & \bm{$R_n$} & \textbf{Au pad size} & \bm{$T_c$} & \bm{$G_{ea}(T_c)$}\\
        & m$\Omega$ & $\mu m \times \mu m$ & mK & pW/K\\
      \hline 
      A & 571  & No Au    & 30 & 8 $\pm$ 1 \\
      B & 572 & 300 $\times$ 300  & 33 & 387 $\pm$ 9  \\
      C & 755 & 600 $\times$ 600  & 35 & 1788 $\pm$ 55 \\
      D & 495  & 900 $\times$ 900  & 35 & 1894 $\pm$ 40 \\
    \end{tabular}
  \end{ruledtabular}
\end{table}

The coefficient $k$ and the exponent $n$ define the thermal conductance $G_{ea}$. 
\begin{equation}
   G_{ea}(T_{e,tes}) ~=~n\cdot k\cdot T_{e,tes}^{n-1}.
     \label{eqn:GvsT}
\end{equation}

\begin{figure}[ht]
\includegraphics[width=0.8\linewidth]{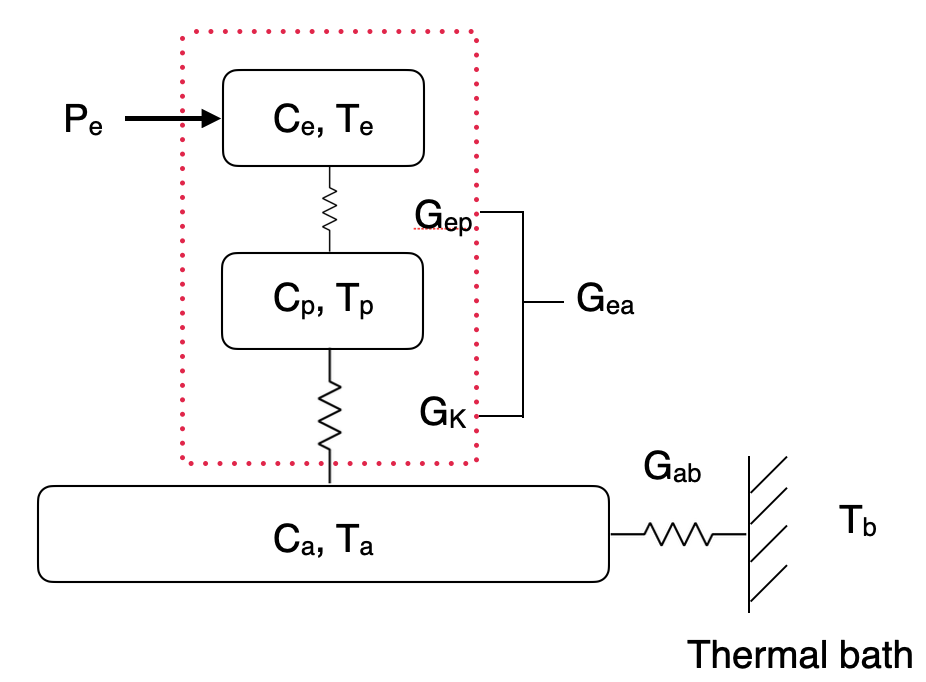}
\caption{\label{fig:thermal_model} Simplified thermal model of the detector. The TES system is shown in the red dotted box. $T_e$ and $C_e$ are the temperature and heat capacity of the electron system of the TES, respectively. The phononic contribution to the heat capacity, $C_p$, of the TES films is extremely small. $T_p \approx T_a$, the temperature of the Si absorber. $T_b$ is the temperature of the thermal bath. Because we use superconducting Al wires, we neglect any thermal connection between the electron system and the thermal bath. $G_{ea}$ and $G_{ab}$ are thermal conductances (see the text for details).}
\end{figure}

A simplified thermal model for our detector is shown in Fig.~\ref{fig:thermal_model}. We assume that the absorber temperature ($T_{a}$) is close to the thermal bath ($T_{b}$).  This assumption is valid only if the thermal conductance between the absorber and the bath, $G_{ab}$, is much larger than that between the TES and the absorber, $G_{ea}$, ensuring $T_{a} \approx T_{b}$. The assumption may not hold for our setup where we are aiming for $G_{ab}$ << $G_{ea}$ to increase the responsitivity for a quantum of energy deposited on the absorber, nonetheless, Eq.~(\ref{eqn:PvsT}) allows us to qualitatively study the effect of Au pads on thermal conductance. 

Eq.~(\ref{eqn:PvsT}) can give an indication of the underlying physics that dominates the conductance of the TESs. The heat flow from the substrate to the sensor is determined by the electron-phonon coupling $G_{ep}$, as well as the Kapitza boundary conductance between the metal film and the substrate $G_{K}$,

\begin{equation}
   \frac{1}{G_{ea}}= \frac{1}{G_{ep}} + \frac{1}{G_{K}}
     \label{eqn:G_sum}
\end{equation}

where $G_{ep} \propto V\cdot T^5$ and $G_{K} \propto A\cdot T^3$ with $V$ and $A$ being the volume and area of the TES, respectively. Since the thickness of the deposited Au is the same for all sensors, $G_{ep} \propto A\cdot T^5$ for the present setup. The value $n$ obtained from the fit to the experimental data indicates the underlying thermal impedance mechanism. All sensors in our device favor a fit with $n=5$. $G_{ea}$ also increases with increasing Au pad area (see Table \ref{tab:table2}). Both observations suggest that our devices' thermal conductance is limited by the electron-phonon coupling and not by the Kapitza impedance. We note that $G_{ea}(T_c)$ does not increase linearly with the Au pad area for very large sizes as expected. We based this on the assumption that we have treated $T_a$ as a constant when the TES power increases. The assumption is not valid for $G_{ab}$ < $G_{ea}$.

We also note that we have ignored the weak-link effects~\cite{smith2013implications} due to which the device's critical current can change for $T_b$ < $T_c$. The weak link effects are severe for very small TESs, but since our TESs are relatively large, we assume that this will not significantly affect our approximation for Eq.~(\ref{eqn:PvsT}).

\section{Heater resolution vs LED resolution}\label{appedixB}
This section qualitatively describes the efficacy of using Poisson statistics of the LED pulses for calibration. In addition, we compare the resolution of the LED pulses with that obtained by the Joule-heater excitation. Unfortunately, since the heater is more susceptible to EMI/RF heating, getting the TES to an appropriate bias point with high sensitivity was nearly impossible. Therefore, we do not quote an absolute energy calibration and $\sigma_{noise}$ in this section since we cannot precisely quantify the amount of residual power and noise in the detector when the heater is connected. For these reasons, we prefer using LED pulses over Joule-heater for detector calibration. 
\begin{figure}[htbp]
\includegraphics[width=\linewidth]{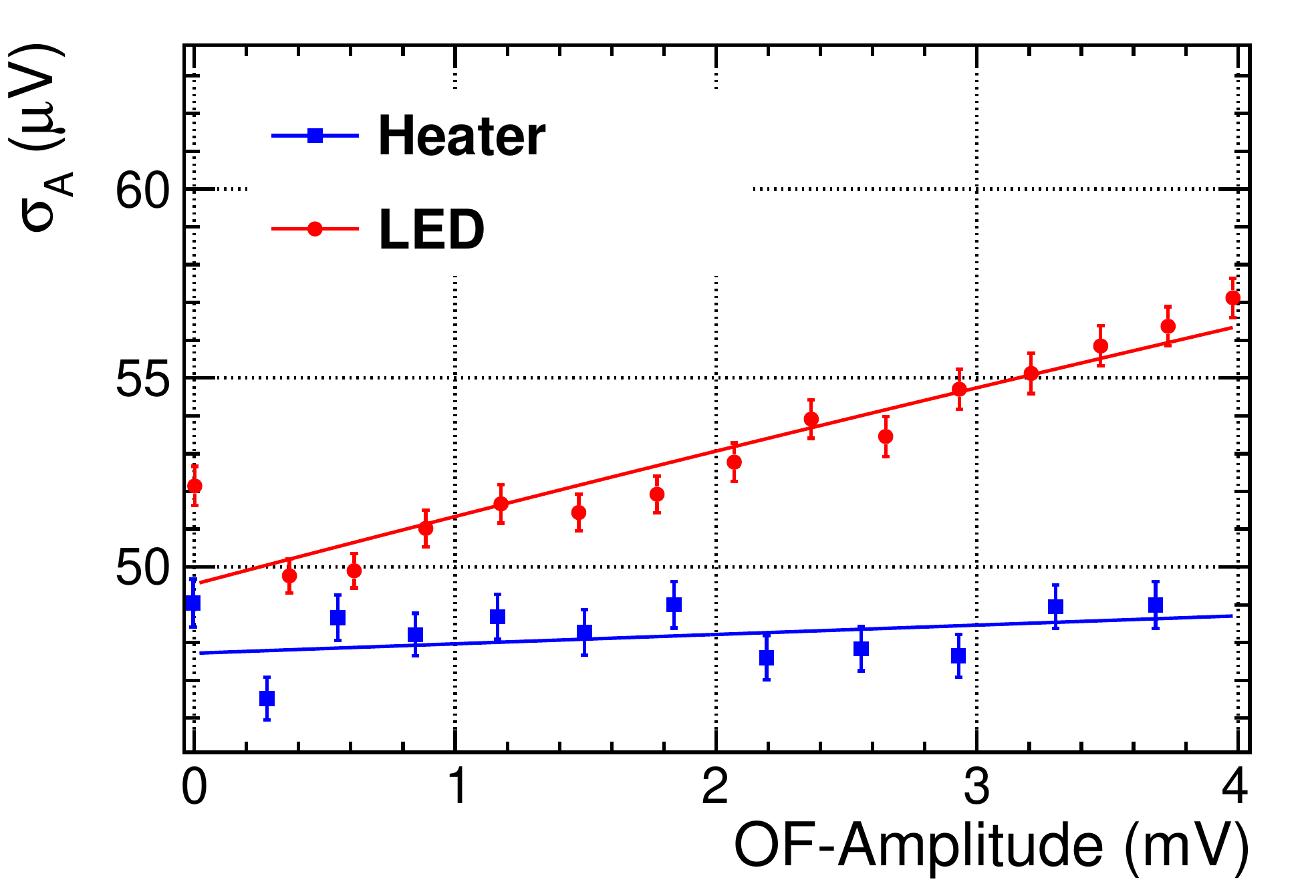}
\caption{\label{fig:heater_led} $\sigma_{A}$ vs. OF-Amplitude for heater and LED excitations. $\sigma_{A}$ for heater excitations are nearly constant, while $\sigma_{A}$ for LED excitations can be fitted to the functional form described in Eq.~\ref{eqn:sigmaA1}. The increase in $\sigma$ for LED excitations is due to Poisson statistics of the number of photons from the LED.}
\end{figure}
Nevertheless, we can still compare the heater and LED excitations as long as we keep the same baseline and temperature for both cases. The heater generates phonon excitations $\mathcal{O}$(few meV) and, unlike LED photons, its $\sigma_{A}$ should not have any amplitude dependence as long as the baseline noise dominates the fluctuation of the number of phonons produced. Since each LED photon has an average energy of 2.1 eV in our case, we expect Poisson statistics to contribute significantly towards resolution for higher-energy LED excitations. As shown in Fig.\ref{fig:heater_led}, this is indeed the case. The heater resolution is constant regardless of the excitation amplitude, while the LED resolution shows an amplitude dependence that can be explained using Eq.~(\ref{eqn:sigmaA1}).

\nocite{*}
\bibliography{aipsamp}

\end{document}